\documentclass[twocolumn,prl,superscriptaddress,showpacs,10pt,aps]{revtex4-1}

\usepackage{bbm}
\usepackage{lmodern,bm}
\usepackage{sarabian}
\usepackage{textcomp}
\usepackage{amsmath}
\usepackage{amsthm}
\usepackage{amsopn}
\usepackage{graphicx}
\usepackage{amsfonts}
\usepackage{amssymb}
\usepackage{mathrsfs}

\usepackage{chapterbib}

\newcommand{\tr}[1]{{\rm tr}\left[#1\right]}
\def\idty{{\bf 1}}

\let\eps\varepsilon

\def\ket #1{\vert#1\rangle}

\let\veps\varepsilon

\newcommand{\sfa}{{\sf A}}
\newcommand{\sfb}{{\sf B}}
\newcommand{\sfc}{{\sf C}}
\newcommand{\sfd}{{\sf D}}

\newcommand{\boa}{\bm{a}}
\newcommand{\bob}{\bm{b}}

\newcommand{\boc}{\bm{c}}
\newcommand{\bod}{\bm{d}}

\newcommand{\bosig}{{\boldsymbol\sigma}}

\newcommand{\conj}{\alpha\bar{\beta}+\bar{\alpha}\beta}

\begin{document}


\title{Direct tests of measurement uncertainty relations: what it takes}

\author{Paul Busch}
\email{paul.busch@york.ac.uk}
\affiliation{University of York, York YO10 5DD, UK} 
\author{Neil Stevens}
\affiliation{University of York, York YO10 5DD, UK} 

\begin{abstract} 

\noindent
The uncertainty principle being a cornerstone of quantum mechanics, it is surprising that in nearly 90 years there have been no direct tests of measurement uncertainty relations. This lacuna was due to the absence of two essential ingredients: appropriate measures of measurement error (and disturbance), and precise formulations of such relations that are {\em universally valid}and {\em directly testable}. We formulate two distinct forms of direct tests, based on different measures of error. We present a prototype protocol for a direct test of measurement uncertainty relations in terms of {\em value deviation errors} (hitherto considered nonfeasible), highlighting the lack of universality of these relations. This shows that the formulation of universal, directly testable measurement uncertainty relations for {\em state-dependent} error measures remains an important open problem. Recent experiments that were claimed to constitute invalidations of Heisenberg's error-disturbance relation, are shown to conform with the spirit of Heisenberg's principle if interpreted as direct tests of measurement uncertainty relations for error measures that quantify {\em distances between observables}.

\end{abstract}

\maketitle

\begin{cbunit}

\emph{Introduction.}\quad Heisenberg's uncertainty principle is arguably one of the most fundamental insights of modern science. With ever-increasing experimental powers of controlling single quantum objects, uncertainty relations are no longer merely of  philosophical significance but must be taken into account in considerations of the limitations to preparing and measuring such systems. Still, it must be noted that there is hitherto no attempt at a direct experimental test of the principle. 

As noted in \cite{BuHeLa07}, there is a small number of publications dedicated to testing the standard {\em preparation} uncertainty relation, but the interpretation of the relevant experiments involves substantial chains of theoretical reasoning, making it impossible to discern which part of quantum mechanics is being tested. In contrast, a {\em direct test} would have to be based on a computation of the measures of uncertainty, or measurement error  or disturbance, under consideration directly from the statistical data obtained for the observables being measured. 

Here we make precise and illustrate the notion of a direct test for {\em measurement uncertainty relations}, that is, trade-off inequalities for approximation errors in joint measurements  (which include error-disturbance relations as a special case, see, e.g., \cite{BuHeLa07}). We discuss the possibility of realizing such direct tests for two proposed quantum generalizations of the classic rms deviation formula for the quantification of measurement errors: (a) rms deviation of values; (b) a distance measure for observables. 

A important practical difference between these two types of error measures is that the former is state-dependent while the latter is state-independent and serves as a figure of merit for the ability of a device to approximate a given observable. 
Recent claims of experimental  violations of Heisenberg's error-disturbance relation (e.g. \cite{Roz12,Erh12,Baek13,White14,Kaneda14}) are found to involve {\em indirect tests} of relations due to Ozawa \cite{Ozawa2004} and Branciard \cite{Branciard2013} that are formulated in terms of error and disturbance measures $\eps,\eta$ akin to rms value deviations. The tests performed so far involve either a weak-value method or the so-called three-state method, both of which provide very indirect determinations of the error measures. The proposed direct test is illustrated in supplementary notes \cite{suppl} with the example of an approximate qubit measurement scheme that was originally designed in \cite{Lund10} to highlight the weak-value method and first realized in \cite{Roz12}.

As elucidated in \cite{BLW13a},  measurement uncertainty relations based on the rms value deviation measure fail to be {\em universal} since the error interpretation of these relations is restricted to a limited class of approximate joint measurements \cite{wrong}. Here we show that this kind of uncertainty relations does lend itself  to a direct test method, one that appears counter-intuitive at first sight and hence was considered infeasible. While this provides a prototype for direct tests of state-dependent measurement uncertainty relations, we find that this method only works for those approximations identified in \cite{BLW13a} for which the error interpretation of these relations is valid.

We therefore conclude that the formulation of universal, directly testable measurement uncertainty relations in terms of state-dependent error measures remains an important open problem, for which error measures other than rms value deviations must be sought (for an interesting recent proposal, see \cite{Renes2014}).

The allegations of violations of Heisenberg's relation have been answered with proofs of new forms of measurement uncertainty relations for state-independent error measures \cite{BLW13b,BLW14a}. The ensuing debate can be clarified by taking note of the distinctions between state-dependent and state-independent error measures \cite{BLW13a} and between direct and indirect tests. We conclude with the observation that rather than refuting Heisenberg's principle, the existing experiments provide direct confirmations of Heisenberg-type measurement uncertainty relations for {\em state-independent}, error and disturbance measures, defined as distances between observables.

\emph{Error as rms value deviation and its direct determination.}\quad
A necessary requirement for a {\em direct test} of a measurement uncertainty relation is that the theoretical values of the measures of error and disturbance used  can be compared with estimates obtained by way of an {\em error analysis} based on the data of the experiment at hand. In the case of the first error measure mentioned above, given as the rms value deviation, this means that an approximate measurement of some observable $A$ is to be performed jointly with a (highly) accurate control measurement of $A$; the statistics thus obtained consists of a distribution of value pairs, for which the rms deviation can be computed and compared with the theoretical value. We spell out conditions under which this procedure becomes meaningful as an error analysis.

We consider the following generic scenario. An observable, represented by selfadjoint operator $A$, is to be measured approximately by a scheme actually measuring some general observable, described by the positive operator valued measure (POVM) $\sfc$. The measurement will generally disturb any other  observable, represented by operator $B$, and distort it into some observable (POVM) $\sfd$. It is known that a  measurement of $\sfc$ followed by an accurate measurement of $B$ constitutes a joint measurement of $\sfc$ and the ``distorted'' observable $\sfd$. 

We first recall the measures of error, $\veps$,  and disturbance, $\eta$, used in the studies 
\cite{Ozawa2004,Roz12,Erh12,Branciard2013,Baek13,White14,Kaneda14}; with respect to the above scenario, these can be expressed as follows: 
\begin{align}
\veps(A)^2&=\iint(x-y)^2{\rm Re}\,\tr{\rho\sfa(dx)\sfc(dy)},\label{eps1}\\
\eta(B)^2&=\iint(x-y)^2{\rm Re}\,\tr{\rho\sfb(dx)\sfd(dy)};\label{eta1}
\end{align}
here $\rho$ is a general density operator of the object and $\sfa,\sfb$ denote the spectral measures of $A,B$.  These expressions are reduce to the classic rms deviation  if $A$ commutes with $\sfc$ and $B$ with $\sfd$; in this case the above equations simplify into
\begin{align}
\veps(A)^2&=\iint(x-y)^2\tr{\rho\sfa(dx)\sfc(dy)},\label{eps2}\\
\eta(B)^2&=\iint(x-y)^2 \tr{\rho\sfb(dx)\sfd(dy)}.\label{eta2}
\end{align}
In this commutative case the quantities $\veps,\eta$ have a proper probabilistic interpretation as the mean values of the squared deviations of the random variables $x,y$ in the probability (bi-)measures defined by $\mu(dx,dy)=\tr{\sfa(dx)\sfc(dy)}$ and
$\nu(dx,dy)=\tr{\sfb(dx)\sfd(dy)}$, respectively. However, as shown in \cite{BLW13a}, if  $A,\sfc$ and $B,\sfd$ do not commute, the quantities $\veps,\eta$ will no longer represent error and disturbance faithfully. In other words, the  interpretation of the inequalities of Ozawa \cite{Ozawa2004} and Branciard \cite{Branciard2013} as error-disturbance relations is limited, in general, to the case of such experiments where the said commutativities are given; they cannot be considered universal.

We focus on the commutative case and consider observables $A,B$ with discrete spectra. We assume that $A$ has values $a_k$, with spectral projections $A_k$, and we consider $\sfc$ to be discrete with values $c_\ell$, where the associated positive operators $C_\ell$ are assumed to commute with the $A_k$. If the measurement of $\sfc$ is preceded by a L\"uders (also known as projective) measurement of $A$, the joint probability for an outcome pair $(a_k,c_\ell)$ is in fact given by 
\begin{align*}
P(A=a_k,\sfc=c_\ell)=\tr{\rho A_kC_\ell}.
\end{align*}
Thus, one can write $\veps(A)$ as a true value-comparison error, testable by preceding the $\sfc$ measurement with a strong (L\"uders) measurement of $A$:
\begin{align}
\veps(A)^2=\sum_{k\ell}(a_k-c_\ell)^2\tr{\rho A_kC_\ell}.\label{Oz-eps}
\end{align}
 
As simple and obvious this procedure appears once it is presented, it was never explicitly stated (to our knowledge), on the gorunds that the initial sharp (or {\em strong}) $A$ measurement  would strongly disturb the state, $\rho$, so that it is not clear whether the measurement of $\sfc$ still can be said to approximately measure $A$ in $\rho$. Instead, it was proposed in \cite{Lund10} to replace the strong $A$ measurement by a so-called weak, as that would hardly disturb the state and the $\sfc$ measurement would still essentially ``see" that state. However, the joint distribution of values of the weak measurement and the $\sfc$ measurement does not {\em directly} render the error quantity $\veps(A)$ as its rms value deviation; instead, a rather complicated reconstruction formula has to be applied to obtain the value of $\veps(A)$ \cite{suppl}. 

On further reflection, it turns out that the effect of the L\"uders measurement is not as disruptive as it appears at first. If $A$ and $\sfc$ commute, it follows that the  $A$ measurement does not disturb the $\sfc$ statistics, that is, the $\sfc$ measurements is still  presented with the same statistics as given by the state $\rho$. The effect of the $A$ measurement is that it feeds the $\sfc$ device with an ensemble of $A$ eigenstates, for which the inaccuracy of $\sfc$ is appropriately quantified by the rms value deviation. The quantity $\eps(A)^2$ is a weighted average of the squared value deviations between the two measurements, and it becomes evident that it encompasses preparation uncertainty in addition to error contributions, due to the state-dependence of the weight factors.

An analogous consideration applies to the disturbance $\eta$.
For an observable  $B$ with discrete values $b_k$ and spectral projections $B_k$ and a distorted $B$ observable $\sfd$ with the same values and positive operators $D_\ell$ (with $\sum_\ell D_\ell=\idty$ and all $D_\ell$ commuting with $B_k$), the expression \eqref{eta2} becomes
\begin{align}
\eta(B)^2=\sum_{k\ell}(b_k-b_\ell)^2\tr{\rho B_kD_\ell}.\label{Oz-eta}
\end{align}

It is possible to give a direct operational implementation of the probability distribution $(k,\ell)\mapsto \tr{\rho B_kD_\ell}$ as follows.
Suppose an approximate measurement of $A$ represented by POVM $\sfc$ is followed by a sharp measurement of observable $B$. This sequential scheme defines a joint measurement of $\sfc$ and some POVM $\sfd$, which is an approximation of $B$. Assume that the disturbance is {\em benign}, in the sense that the $D_\ell$ commute with the $B_k$, which occurs, for example, when $\sfd$ is a smearing of $B$ by means of a stochastic matrix $(\lambda_{\ell m})$, i.e., $D_\ell=\sum_m\lambda_{\ell m}B_m$. Now assume that the measurement of $\sfc$ is preceded by a projective measurement of $B$. It follows that the operational joint probabilities are
\begin{align*}
P(B_i=b_k,B_f=b_\ell,\sfc=c_n)&=\tr{\mathcal{I}^\sfc_n(B_k\rho B_k) B_\ell}\\
&=\tr{B_k\rho B_k{(\mathcal{I}^\sfc_n)}^*(B_\ell)}.
\end{align*}
Here $n\mapsto \mathcal{I}^\sfc_n$ denotes the instrument associated with $\sfc$, giving the state change conditional on the outcome $n$,
and ${(\mathcal{I}^\sfc_n)}^*$ is the dual of the operation $\mathcal{I}^\sfc_n$. Disregarding the outcomes of the $\sfc$ measurement and noting that $D_\ell=\sum_n{(\mathcal{I}^\sfc_n)}^*(B_\ell)$, we obtain the marginal probability
\begin{align*}
P(B_i=b_k,B_f=b_\ell)&=\tr{B_k\rho B_kD_\ell}\\
&=\tr{\rho B_k D_\ell},
\end{align*}
since $B_k$ commutes with $D_\ell$. This is the joint probability of obtaining  values $b_i$ and $b_\ell$ in accurate $B$ measurements preceding and succeeding a measurement of $\sfc$.
Therefore,
\begin{align*}
\eta(B)^2=\sum_{k,\ell}(b_k-b_\ell)^2P(B_i=b_k,B_f=b_\ell).
\end{align*}

Again, it may be (and was) thought that the effect of the initial sharp $B$ measurement invalidates the intermediate measurement of $\sfc$ as an approximation of $A$. However, all the initial measurement does is that it feeds the $\sfc$ device with  $B$ eigenstate, in addition to indicating the corresponding $B$ eigenvalue. The quantity $\eta(B)^2$ is thus seen to represent the squared deviation of the values of the initial and final $B$ measurements, averaged over the ensemble of $B$ eigenstates that $\sfc$ ``sees'', where the weights depend on the input state $\rho$. This analysis also highlights the fact that $\eta(B)$ encompasses preparation uncertainty in addition to disturbance contributions.

 It is evident from the above constructions that  these direct test procedures are not applicable if the observable pairs $A, \sfc$ and $B,\sfd$ do not commute since in that case their rms value deviations do not yield $\eps(A),\eta(B)$,  However, within the realm of commuting approximations, our schemes are proof of principle that direct tests of measurement uncertainty relations can be performed. In contrast,  the experiments carried out so far are based on indirect determinations of $\eps(A)$ and $\eta(B)$. Since these quantities fail to  represent error and disturbance in general, the inequalities of Ozawa and Branciard cannot be said to be universal measurement
 uncertainty relations \cite{BLW13a}.
 
Here we briefly recall the example proposed by Lund and Wiseman, a model experimental determination of $\eta(B)$ for qubit observables, that does fall into the class of schemes where the commutativity of  $B$ with the distorted observable $\sfd$ is given. 
In this model (Fig. \ref{Fig1}) an initial approximate (or weak)  measurement of the qubit observable $B=X$  \cite{Pauli} is done, with strength $2\gamma^2-1$ (where $\gamma$ appears as a parameter in the initial state of the weak measurement probe, $\gamma\ket0+(1-\gamma^2)^{1/2}\ket1$, hence $\gamma^2\le 1$). This is then followed by an approximate measurement of $Z$ on the resulting state, with strength $\cos{2\theta}$. Finally there is an accurate $X$ measurement (denoted $X_f$). The initial and final $X$ measurements are intended to provide information about the disturbance of $X$ by the approximate $Z$ measurement. The probe and measurement system performing the first $X$ measurement and the approximate $Z$ measurement are again qubit observables, and their readout observables are $Z_p$ and $Z_m$, respectively.

The authors of \cite{Lund10} and \cite{Roz12,White14} focus on the ``weak'' limit, $\gamma^2\approx 1/2$. The underlying intuition is that thereby the approximate $Z$ measurement still receives the practially undisturbed input state $\rho$. Alternatively, one could disregard this desideratum and set the strength parameter to its maximum value (obtained for $\gamma^2=1$); this renders the initial
 measurement a ``strong'', or sharp measurement of $X$, feeding the approximate $Z$ measurement with $X$ eigenstates.
 
In the Supplement \cite{suppl} we give a detailed analysis of this scheme to highlight the contrast between the very indirect weak method and the direct test emerging in the strong limit.

\begin{figure}[ht]
\centering
  \includegraphics[width=8.5cm]{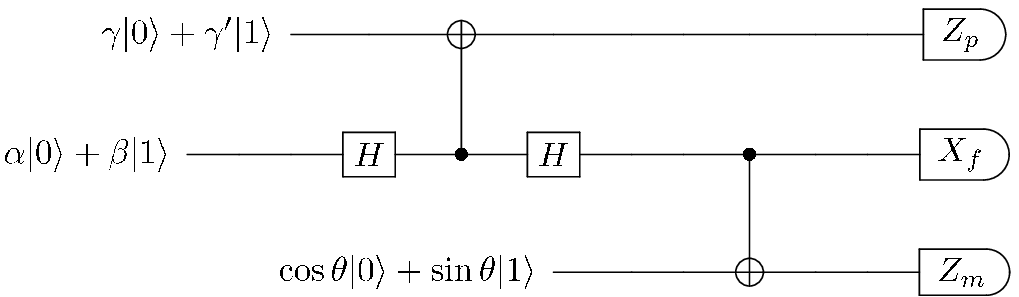}
	\caption{Model implementation of a determination of $\eta(X)$. The top and bottom wires represent the probe and measuring system while the middle wire corresponds to the observed qubit. As shown in the Supplement, the value of $\eta(X)$ can be extracted from the joint distribution of the initial and final $X$ measurements, obtained by reading the outputs $Z_p$ and $X_f$.}\label{Fig1}
\end{figure}

{\em Error as distance between observables.}\quad

The state-independent error and disturbance measures used in \cite{BLW13b,BLW14a} is based on a comparison of the statistics of the two observables to be compared, such as an observable $A$ being approximated by $\sfc$, or $B$ being distorted into $\sfd$. We briefly recall the definition of these measures as distances.

 For any two (discrete) probability distributions $p:x_k\mapsto p_k$, $q:y_\ell\mapsto q_\ell$,  a {\it coupling} is defined to be a joint probability distribution $\gamma:(x_k,y_\ell)\mapsto \gamma_{k\ell}$  with  $p$ and $q$ as its Cartesian marginals.
The set of couplings between $p$ and $q$ will be denoted $\Gamma(p,q)$. Then, the (Wasserstein) 2-distance \cite{Villani} of $p$ and $q$ is defined as
\begin{equation*}
  \mathcal{D}_2(p,q) =  
  \inf_{\gamma\in\Gamma(p,q)}\left(\sum (x_k-y_\ell)^2\,\gamma_{k\ell} \right)^{\frac12}
\end{equation*}
This is a distance between probability measures due to the choice of the minimizing joint probability. 

The \emph{(Wasserstein) $2$-distance} between two observables, say $a_k\mapsto A_k$ and $c_\ell\mapsto C_\ell$, is then defined as follows, using the notation $p_\rho^A$, $p_\rho^{\sfc}$ for the probability distributions of $A,\sfc$ with respect to the state $\rho$:
\begin{equation*}
\Delta_2(A,\sfc):=\sup_{\rho}\mathcal{D}_2(p_\rho^A,p_\rho^{\sfc}).
\end{equation*}
This distance between the observable can be determined from the statistics $p_\rho^A,p_\rho^{\sfc}$, obtained in separate runs of $A$ and $\sfc$ measurements on different ensembles of systems prepared in the same state $\rho$. The method does not depend on whether or not $A$ and $\sfc$ commute. In the commutative case, since $\eps(A)$ is obtained from a particular coupling of the distributions of $A$ and $\sfc$, it is always true that 
$
\eps(A)\ge\mathcal{D}_2(p_\rho^A,p_\rho^{\sfc}),
$
so that the former is always an upper bound estimate of the latter (metric) quantity. 

The relevant error analysis for a measurement of an observable $\sfc$ as an approximation of $A$ is simply the comparison of these statistics.  Since the 2-distance between observables is defined solely in terms of their probabilities, which are obtained from their measurement statistics, it follows that any uncertainty relation formulated in terms of this distance is automatically directly testable. The statistics of $A$ and $\sfc$ are obtained by independent runs of measurements, so that there is no restriction to the class of approximations $\sfc$ to an observable $A$. In other words, such uncertainty relations are universal.

Examples of measurement uncertainty relations for distances between observables are known for position and momentum \cite{BLW13b} and qubit observables \cite{BLW14a}. Surprisingly, the existing experimental tests of the Ozawa and Branciard inequalities, while failing as direct tests for rms value deviations, can be reinterpreted as direct tests and confirmations of the qubit inequality for distances found in \cite{BLW14a}, as we show next.

\emph{Testing qubit uncertainty relations.}\quad
Interestingly, $\veps(A)$ and $\eta(B)$ become entirely state-independent in the case of qubit observables
within a class of approximating observables that are optimal in the sense of the 2-distance. The cited experimental
tests of the Ozawa and Branciard inequalities use such approximators, but not much is made of the curious state-independence of $\eps,\eta$. In fact, these quantities are then directly related to the 2-distance measures $\Delta(A,\sfc),\Delta(B,\sfd)$.
 This explains why  the qubit experiments utilizing either the three-state method or the weak value method can serve as {\em direct test} of any trade-off for these metric error and disturbance measures. 

As was shown in \cite{BLW13a}, the existing experiments realize approximating observables $\sfc,\sfd$ of the form $C_\pm=\frac12(\idty\pm\boc\cdot\bosig)$ and $D_\pm=\frac12(\idty\pm\bod\cdot\bosig)$, where the target observables are $A=\boa\cdot\bosig$ and $B=\bob\cdot\bosig$, respectively. (Here we use the Bloch vector representation of operators in the two-dimensional Hilbert space for qubits, so that $A,B$ are associated with unit vectors $\boa,\bob$, etc., and $\bosig=(\sigma_1,\sigma_2,\sigma_3)$ represents the Pauli matrix triple.) Observables $\sfc,\sfd$ of this kind are known to give optimal approximations, in the sense that for any general approximating observable one can always find a better approximator (with smaller distances $\Delta(A,\sfc),\Delta(B,\sfd)$) from this class \cite{BuHe08}.  

The distance can be evaluated as \cite{BLW14a}
\[
\Delta(\sfa,\sfc)^2=2\|\boa-\boc\|.
\]
The quantity $\veps(A)$ is directly related to this distance:
\[
\veps(A)^2\,=\, 1-\|\boc\|^2+\tfrac14\Delta(\sfa,\sfc)^4\,\le\,\Delta(\sfa,\sfc)^2
\]
(confirming its state-independence in this case).
In the Vienna experiment \cite{Erh12}, the approximators are misaligned sharp observables ($\|\boc\|=1$), giving $\veps(A)=\frac12 \Delta(\sfa,\sfc)^2$.
In the Toronto experiment \cite{Roz12}, they are smearings of the target with $\boc=\lambda\boa$, hence commuting, and one has $\veps(A)=\Delta(\sfa,\sfc)$.

$\Delta(\sfa,\sfc)$ is directly obtained from the statistics of the $A$ and $\sfc$ measurements for sufficiently many states since this
yields estimates of $\boa,\boc$. Alternatively, this number can be calculated using the value of $\eps(A)$ obtained in the experiments mentioned. It follows that these experiments serve in fact as direct tests of a universal error-disturbance relation for worst-case errors and disturbances, namely Branciard's inequality in the form \cite[Eq. (12)]{Branciard2013}, evaluated for the
observables $A=Z=\sigma_3,B=X=\sigma_1$ and a $Y(=\sigma_2)$ eigenstate. Using the scaling $\veps(Z)^2=2d_Z$, $\eta(X)^2=2d_X$, this inequality reads simply
\[
(d_Z-1)^2+(d_X-1)^2\le 1,
\]
with values of interest being $d_Z,d_X\le 1$. In the case of commuting approximators this strengthens the inequality $d_Z+d_X\ge 2-\sqrt2$ obtained in
\cite{BuHe08,BLW14a}, with $d_Z,d_X$ now equal to $\|\boa-\boc\|,\|\bob-\bod\|$. Rather than being violations of Heisenberg's principle, the experiments thus confirm inequalities that are very much in the spirit of Heisenberg's uncertainty ideas \cite{violation}.

To conclude, we have formulated the concept of a direct test of measurement uncertainty relations for error measures given either as rms value deviations or as distances between observables. We have shown that such tests can be realized in principle, but the first error measure is of limited applicability and hence does not give rise to universal uncertainty relations. It remains an important open problem to find {\em state-specific} error measures that yield {\em universal} and directly testable error-disturbance relations with nontrivial trade-off bounds.

\vbox{
\section*{Acknowledgements}  N.S. gratefully acknowledges support through the award of an Annie Currie Williamson PhD Bursary at the University of York. Thanks are due to Pekka Lahti for helpful critical comments on various draft versions of this work.
}

\end{cbunit}

\begin{cbunit}

\onecolumngrid

\section*{Supplement\\ Determining measurement errors: weak vs. strong measurement method}

The experimental setup proposed by Lund and Wiseman \cite{Lund10} consists of a three-qubit system, the object in initial state $\alpha\ket{0} + \beta\ket{1}$, a ``weak measurement probe ($p$)" initially in state $\gamma\ket{0} + \gamma'\ket{1}$, and the apparatus $m$ with initial state $\cos{\theta}\ket{0} + \sin{\theta}\ket{1}$, all in their respective 2-dimensional  Hilbert spaces $\mathscr{H}$, $\mathscr{H}_p$ and$\mathscr{H}_m$, respectively.

In the scenario when the disturbance measure for the observable $X$ is to be determined \cite{Pauli}, the initial approximate $X$ measurement is enacted by first applying a Hadamard gate on the object system $\mathscr{H}$, followed by a $C_{NOT}$ gate acting on $\mathscr{H}_p$, controlled on $\mathscr{H}$, and finally with another Hadamard gate performed on $\mathscr{H}$. This is followed by the device whose disturbance is being measured, wherein a $C_{NOT}$ gate acts on $\mathscr{H}_m$, again controlled on $\mathscr{H}$. Sharp $Z$ measurements are then performed on $\mathscr{H}_p$ and $\mathscr{H}_m$, (denoted $Z_p$ and $Z_m$ respectively), along with a sharp $X$ measurement ($X_f$) on $\mathscr{H}$ (Fig.~\ref{Fig1}).

\begin{figure}[ht]
\centering
  \includegraphics[width=12cm]{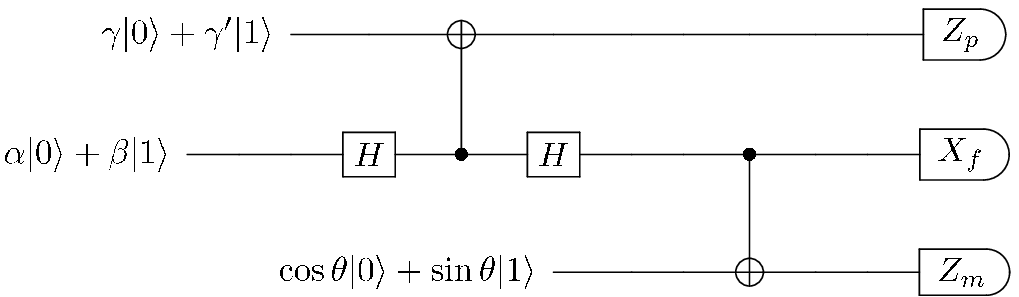}
	\caption{Model implementation of a determination of $\eta(X)$. The top and bottom wires represent the probe and measuring system while the middle wire corresponds to the observed qubit. As shown in the text, the value of $\eta(X)$ can be extracted from the joint distribution of the initial and final $X$ measurements, obtained by reading the outputs $Z_p$ and $X_f$.}\label{Fig1}
\end{figure}

The scheme thus realises a joint (sequential) measurement of three $\pm 1$ valued observables, with probabilities
\begin{equation*}
P_{k,\ell,n}:=P(Z_p=k,X_f=\ell,Z_m=n),\quad k,\ell,n\in\{+,-\},
\end{equation*}
which are determined next together the associated POVMs (see also \cite{Mikko}).

The state of the object and weak probe combined $|\psi_1\rangle$, after the the initial interaction is then given by:
\begin{align*}
|\psi_1\rangle =\, &(\mathbb{I}\otimes H)C_{NOT}(\mathbb{I}\otimes H)(\gamma\ket{0} + \gamma'\ket{1})\otimes(\alpha\ket{0} + \beta\ket{1}) \\
=\, &\frac{1}{\sqrt{2}}(\mathbb{I}\otimes H)C_{NOT}(\gamma\ket{0} + \gamma'\ket{1})\otimes\left( (\alpha + \beta)\ket{0} + (\alpha - \beta)\ket{1}\right) \\
=\, &\frac{1}{\sqrt{2}}(\mathbb{I}\otimes H)\left[(\gamma\ket{0} + \gamma'\ket{1})\otimes(\alpha + \beta)\ket{0} + (\gamma'\ket{0} + \gamma\ket{1})\otimes(\alpha - \beta)\ket{1}\right] \\
=\, &\frac{1}{2}[(\gamma(\alpha + \beta)+\gamma'(\alpha - \beta))\ket{0}\otimes\ket{0} + (\gamma'(\alpha + \beta)+\gamma(\alpha - \beta))\ket{1}\otimes\ket{0} \\
&\quad + (\gamma(\alpha + \beta)-\gamma'(\alpha - \beta))\ket{0}\otimes\ket{1} + (\gamma'(\alpha + \beta)-\gamma(\alpha - \beta))\ket{1}\otimes\ket{1}] \\
=\, &|p_0\rangle\otimes\ket{0} + |p_1\rangle\otimes\ket{1},
\end{align*}
where
\begin{align*}
|p_0\rangle =\, &\frac{1}{2}(\gamma(\alpha + \beta)+\gamma'(\alpha - \beta))\ket{0} + (\gamma'(\alpha + \beta)+\gamma(\alpha - \beta))\ket{1} \\
|p_1\rangle =\, & \frac{1}{2} (\gamma(\alpha + \beta)-\gamma'(\alpha - \beta))\ket{0} + (\gamma'(\alpha + \beta)-\gamma(\alpha - \beta))\ket{1}.
\end{align*}
The state of the whole system after the measuring device, $|\psi_f\rangle$ is then
\begin{align*}
|\psi_f\rangle =\, &(\mathbb{I}\otimes C_{NOT})(|p_0\rangle\otimes\ket{0} + |p_1\rangle\ket{1})\otimes(\cos{\theta}\ket{0} + \sin{\theta}\ket{1}) \\
=\, &|p_0\rangle\otimes\ket{0}\otimes(\cos{\theta}\ket{0} + \sin{\theta}\ket{1}) + |p_1\rangle\otimes\ket{1}\otimes(\sin{\theta}\ket{0} + \cos{\theta}\ket{1}) \\
=\, &|p_0\rangle\otimes\ket{0}\otimes|m_0\rangle + |p_1\rangle\otimes\ket{1}\otimes|m_1\rangle,
\end{align*}
with
\begin{align*}
|m_0\rangle =\, &\cos{\theta}\ket{0} + \sin{\theta}\ket{1} \\
|m_1\rangle =\, &\sin{\theta}\ket{0} + \cos{\theta}\ket{1}.
\end{align*}
Now writing $|+\rangle = \frac{1}{\sqrt{2}}(\ket{0}+\ket{1})$ and $|-\rangle = \frac{1}{\sqrt{2}}(\ket{0}-\ket{1})$, the eigenstates of $X$, we have
\begin{align*}
|\psi_f\rangle =\, &\frac{1}{\sqrt{2}}[|p_0\rangle\otimes|+\rangle\otimes|m_0\rangle + |p_1\rangle\otimes|+\rangle\otimes|m_1\rangle + |p_0\rangle\otimes|-\rangle\otimes|m_0\rangle - |p_1\rangle\otimes|-\rangle\otimes|m_1\rangle] \\
=\, &\frac{1}{2\sqrt{2}}[([\gamma(\alpha+\beta)+\gamma'(\alpha-\beta)]\cos{\theta} + [\gamma(\alpha+\beta)-\gamma'(\alpha-\beta)]\sin{\theta})\ket{0}\otimes|+\rangle\otimes\ket{0} \\
&\qquad+([\gamma(\alpha+\beta)+\gamma'(\alpha-\beta)]\sin{\theta} + [\gamma(\alpha+\beta)-\gamma'(\alpha-\beta)]\cos{\theta})\ket{0}\otimes|+\rangle\otimes\ket{1} \\
&\qquad+([\gamma'(\alpha+\beta)+\gamma(\alpha-\beta)]\cos{\theta} + [\gamma'(\alpha+\beta)-\gamma(\alpha-\beta)]\sin{\theta})\ket{1}\otimes|+\rangle\otimes\ket{0} \\
&\qquad+([\gamma'(\alpha+\beta)+\gamma(\alpha-\beta)]\sin{\theta} + [\gamma'(\alpha+\beta)-\gamma(\alpha-\beta)]\cos{\theta})\ket{1}\otimes|+\rangle\otimes\ket{1} \\
&\qquad+([\gamma(\alpha+\beta)+\gamma'(\alpha-\beta)]\cos{\theta} - [\gamma(\alpha+\beta)-\gamma'(\alpha-\beta)]\sin{\theta})\ket{0}\otimes|-\rangle\otimes\ket{0} \\
&\qquad+([\gamma(\alpha+\beta)+\gamma'(\alpha-\beta)]\sin{\theta} - [\gamma(\alpha+\beta)-\gamma'(\alpha-\beta)]\cos{\theta})\ket{0}\otimes|-\rangle\otimes\ket{1} \\
&\qquad+([\gamma'(\alpha+\beta)+\gamma(\alpha-\beta)]\cos{\theta} - [\gamma'(\alpha+\beta)-\gamma(\alpha-\beta)]\sin{\theta})\ket{1}\otimes|-\rangle\otimes\ket{0} \\
&\qquad+([\gamma'(\alpha+\beta)+\gamma(\alpha-\beta)]\sin{\theta} - [\gamma'(\alpha+\beta)-\gamma(\alpha-\beta)]\cos{\theta})\ket{1}\otimes|-\rangle\otimes\ket{1}].
\end{align*}
From here the probabilities of the various outcomes can be read off; writing, say $P_{+-+}$ for the probability $P(Z_p=+1, X_f=-1, Z_m=+1)$, we have:
\begin{align*}
8P_{+++} =\,& 1+(2\gamma^2-1)(\conj)+\sin({2\theta})[(2\gamma^2-1)+(\conj)]
+2\gamma\gamma'(|\alpha|^2-|\beta|^2)\cos(2{\theta})\\
8P_{++-} =\,& 1+(2\gamma^2-1)(\conj)+\sin({2\theta})[(2\gamma^2-1)+(\conj)]
 -2\gamma\gamma'(|\alpha|^2-|\beta|^2)\cos(2{\theta})\\
8P_{-++} =\,& 1+(1-2\gamma^2)(\conj)+\sin({2\theta})[(1-2\gamma^2)+(\conj)]
+2\gamma\gamma'(|\alpha|^2-|\beta|^2)\cos(2{\theta})\\
8P_{-+-} =\,& 1+(1-2\gamma^2)(\conj)+\sin({2\theta})[(1-2\gamma^2)+(\conj)]
 -2\gamma\gamma'(|\alpha|^2-|\beta|^2)\cos(2{\theta})\\
8P_{+-+} =\,& 1+(2\gamma^2-1)(\conj)-\sin({2\theta})[(2\gamma^2-1)+(\conj)]
+2\gamma\gamma'(|\alpha|^2-|\beta|^2)\cos(2{\theta})\\
8P_{+--} =\,& 1+(2\gamma^2-1)(\conj)-\sin({2\theta})[(2\gamma^2-1)+(\conj)]
 -2\gamma\gamma'(|\alpha|^2-|\beta|^2)\cos(2{\theta})\\
8P_{--+} =\,& 1+(1-2\gamma^2)(\conj)-\sin({2\theta})[(1-2\gamma^2)+(\conj)]
+2\gamma\gamma'(|\alpha|^2-|\beta|^2)\cos(2{\theta})\\
8P_{---} =\,& 1+(1-2\gamma^2)(\conj)-\sin({2\theta})[(1-2\gamma^2)+(\conj)]
 -2\gamma\gamma'(|\alpha|^2-|\beta|^2)\cos(2{\theta}).
\end{align*}
This gives the respective 8-outcome POVM with positive operators $E_{k\ell m}$ on the target system:
\begin{align*}
8E_{+++} =\,& (1+\sin({2\theta})(2\gamma^2-1))\mathbb{I}+(2\gamma^2-1+\sin({2\theta}))X
+2\gamma\gamma'\cos(2{\theta})Z\\
8E_{++-} =\,& (1+\sin({2\theta})(2\gamma^2-1))\mathbb{I}+(2\gamma^2-1+\sin({2\theta}))X
-2\gamma\gamma'\cos(2{\theta})Z\\
8E_{-++} =\,& (1+\sin({2\theta})(1-2\gamma^2))\mathbb{I}+(1-2\gamma^2+\sin({2\theta}))X
+2\gamma\gamma'\cos(2{\theta})Z\\
8E_{-+-} =\,& (1+\sin({2\theta})(1-2\gamma^2))\mathbb{I}+(1-2\gamma^2+\sin({2\theta}))X
-2\gamma\gamma'\cos(2{\theta})Z\\
8E_{+-+} =\,& (1-\sin({2\theta})(2\gamma^2-1))\mathbb{I}+(2\gamma^2-1-\sin({2\theta}))X
+2\gamma\gamma'\cos(2{\theta})Z\\
8E_{+--} =\,& (1-\sin({2\theta})(2\gamma^2-1))\mathbb{I}+(2\gamma^2-1-\sin({2\theta}))X
-2\gamma\gamma'\cos(2{\theta})Z\\
8E_{--+} =\,& (1-\sin({2\theta})(1-2\gamma^2))\mathbb{I}+(1-2\gamma^2-\sin({2\theta}))X
+2\gamma\gamma'\cos(2{\theta})Z\\
8E_{---} =\,& (1-\sin({2\theta})(1-2\gamma^2))\mathbb{I}+(1-2\gamma^2-\sin({2\theta}))X
-2\gamma\gamma'\cos(2{\theta})Z.
\end{align*}
From here we can read off the actual (marginal) 2-outcome POVMs that are being measured on the system at the three stages. Firstly the $Z_p$ measurement
defines the positive operators $P_k=\sum_{\ell m}E_{k\ell m}$ representing the initial weak $X$ measurement:
\begin{align*}
P_+ =\,& \tfrac12\bigl[\,\mathbb{I}+(2\gamma^2-1)X\,\bigr] \\
P_- =\,& \tfrac12\bigl[\, \mathbb{I}-(2\gamma^2-1)X\,\bigr],
\end{align*}
the final sharp $X_f$ corresponds to measuring the POVM $D_{\ell}=\sum_{km}E_{k\ell m}$:
\begin{align*}
D_+ =\,& \tfrac12\bigl[\, \mathbb{I}+\sin({2\theta})X\,\bigr]\\
D_- =\,& \tfrac12\bigl[\, \mathbb{I}-\sin({2\theta})X\,\bigr],
\end{align*}
and the observable actually being measured by the measurement device whose disturbance power is being assessed is $C_m=\sum_{k\ell}E_{k\ell m}$:
\begin{align*}
C_+ =\,& \tfrac12\bigl[\, \mathbb{I}+2\gamma\gamma'\cos(2{\theta})Z\,\bigr]\\
C_- =\,& \tfrac12\bigl[\, \mathbb{I}-2\gamma\gamma'\cos(2{\theta})Z\,\bigr]
\end{align*}

We also note down the POVM, $F_{k\ell}=\sum_m E_{k\ell m}$,  representing the joint measurement of the initial weak $X$ observable and the final $X_f$ measurement, which is used to calculate the disturbance quantity:
\begin{align*}
F_{++} =\,&\tfrac14\bigl[\, (1+\sin({2\theta})(2\gamma^2-1))\mathbb{I}+(2\gamma^2-1+\sin({2\theta}))X\,\bigr]\\
F_{-+} =\,& \tfrac14\bigl[\,  (1-\sin({2\theta})(2\gamma^2-1))\mathbb{I}-(2\gamma^2-1-\sin({2\theta}))X\,\bigr]\\
F_{+-} =\,&\tfrac14\bigl[\,  (1-\sin({2\theta})(2\gamma^2-1))\mathbb{I}+(2\gamma^2-1-\sin({2\theta}))X\,\bigr]\\
F_{--} =\,&\tfrac14\bigl[\,  (1+\sin({2\theta})(2\gamma^2-1))\mathbb{I}-(2\gamma^2-1+\sin({2\theta}))X\,\bigr].
\end{align*}
The associated operational joint probabilities in the state $\alpha\ket{0} + \beta\ket{1}$ are (putting $\langle X\rangle=\conj$):
\begin{align*}
P(Z_p=+1,X_f=+1)=\,& \tfrac14\bigl[\, (1+\sin({2\theta})(2\gamma^2-1))+(2\gamma^2-1+\sin({2\theta}))\langle X\rangle\,\bigr]   \\
P(Z_p=-1,X_f=+1)=\,&    \tfrac14\bigl[\,  (1-\sin({2\theta})(2\gamma^2-1))-(2\gamma^2-1-\sin({2\theta}))\langle X\rangle\,\bigr]  \\
P(Z_p=+1,X_f=-1)=\,& \tfrac14\bigl[\,  (1-\sin({2\theta})(2\gamma^2-1))+(2\gamma^2-1-\sin({2\theta}))\langle X\rangle\,\bigr]   \\
P(Z_p=-1,X_f=-1)=\,&  \tfrac14\bigl[\,  (1+\sin({2\theta})(2\gamma^2-1))-(2\gamma^2-1+\sin({2\theta}))\langle X\rangle\,\bigr]  .
\end{align*}
With these expressions it is straightforward to evaluate the expression given by Lund and Wiseman,
\begin{equation}\label{eta-wv}
\eta(X)^2 = \sum_{\delta x} (\delta x)^2 P_{WV}(\delta x)
\end{equation}
 for the ``weak-valued probabilities'' $P_{WV}(\delta x)=\sum_{k,\ell:x_\ell=x_k+\delta x}P_{WV}(x_k,x_\ell)$)
used to determine $\eta(X)^2$: 
\begin{align*}
2P_{WV}(\delta X=\pm2) &= 2P_{WV}(X_i=\mp1|X_f=\pm1)P(X_f=\pm1)\nonumber \\
&= P(Z_p=1,X_f=\pm1)+P(Z_p=-1,X_f=\pm1)  \mp \frac{P(Z_p=1,X_f=\pm1)-P(Z_p=-1,X_f=\pm1)}{2\gamma^2-1}.\nonumber
\end{align*}
The last expression, which can be directly evaluated using the above probabilities, is to be compared with the weak-valued probability on the left hand side:
\begin{align*}
P_{WV}(\delta X=\pm2)&=P_{WV}(X_i=\mp1,X_f=\pm1)=\left\langle \frac12(\mathbb{I}\mp X)\frac 12\bigl(\mathbb{I}\pm \sin(2\theta)X\bigr)\right\rangle\\
&=\tfrac12\bigl(1-\sin(2\theta)\bigr)\tfrac12\bigl(1\mp\langle X\rangle\bigr).
\end{align*}
We observe that these expressions for the weak-valued joint probabilities do not coincide with the operational probabilities,
$P(Z_p=\mp1,X_f=\pm1)$, {\em except} in the strong measurement case, $\gamma=1$. Now we can evaluate $\eta(X)^2$ as given in 
Eq.~\eqref{eta-wv}:
\[
\eta(X)^2=\bigl(1-\sin(2\theta)\bigr)\bigl(1-\langle X\rangle\bigr)+\bigl(1-\sin(2\theta)\bigr)\bigl(1+\langle X\rangle\bigr)
=2\left(1-\sin(2\theta)\right)=2(\cos\theta-\sin\theta)^2.
\]

Finally we verify the strong measurement realization of $\eta(X)^2$.
\[
4P(Z_p=+1,X_f=-1)+4P(Z_p=-1,X_f=+1)=\, 2-2\sin({2\theta})(2\gamma^2-1).
\]
Note that this is already state-independent. On putting $\gamma=1$, we finally obtain
\[
4P(Z_p=+1,X_f=-1)+4P(Z_p=-1,X_f=+1)=\, 2-2\sin({2\theta})=\eta(X)^2.
\]

\end{cbunit}


\begin{thebibliography}{9}
\bibitem{wrong} Specifically, it was shown in \cite{BLW13a} that the quantity $\eps$ can be zero although the approximation is evidently bad in the given state. This can even happen for both ``errors'' in an approximate joint measurement, rendering a ``tight'' instance, 0=0, of the Ozawa and Branicard inequalities, while the ``true'' accuracy is definitely not perfect. That something was wrong with the quantity $\veps$ was noted by Branciard \cite{Branciard2013}, who also observed that $\veps$ can be zero while the approximation does not appear to be perfect. He considered this an artefact of the choice of measure and remarked that it would be interesting to consider alternative measures.
\bibitem{Pauli} The observables $X,Y,Z$ are identical to the Pauli operators $\sigma_x,\sigma_y,\sigma_z$.
\bibitem{violation} The claim of a violation of Heisenberg's error-disturbance relation was based on an untenable attribution to Heisenberg of the inequality $\eps(A)\eta(B)\ge\frac12|\langle[A,B]\rangle_\rho|$, which was known to be wrong in general.
\end{thebibliography}
\end{document}